# About some applications of Kolmogorov equations to the simulation of financial institutions activity

Mikhail I. Rumyantsev

Zakhidnodonbaskiy private institute of economics and management, Pavlograd, Ukraine
E-mail: renixa-1959@mail.ru

***Abstract***
*The goal of this article is to describe the concepts of system dynamics and its applications to the simulation modeling of financial institutions daily activity. The hybrid method of the re-engineering of banking business processes based upon combination of system dynamics, queuing theory and tools of ordinary differential equations (Kolmogorov equations) is offered.*

**1. Introduction**

The presence of material, financial and social components in the structure of economical systems and processes requires an application of different tools in accordance with a level of simulation model. As the practice demonstrates, the manufacturing and technological models (frequently considered as systems of queuing) can be adequately modelled by means of discrete-eventual simulation systems such as GPSS. The financial models fit well into the framework of system dynamics; the multi-agent approach can be useful to the simulation modeling of labour relations. Moreover, for a long time, explorers had differing views on the creation of simulation models of economical processes and used miscellaneous approaches to the development of the related software. Many of these views and approaches have found the reflection on pages of the university tutorials (for example [6]).

Of all methods of computer simulation actively used today, our special focus is on system dynamics. The foundation of this theory was laid in the 1950s by J. Forrester, who published a famous article about industrial dynamics in 1958. System dynamics was offered as a tool for the research of the behavior of industrial systems by means of detecting information feedbacks. The purpose of this procedure is to study the interaction of the structure, amplifications in business policies, and delays in decision making and to evaluate the influence of these factors on the efficiency of a firm and on its business operations. To quote J. Forrester: "System dynamics combines the theory, methods, and philosophy needed to analyze the behavior of systems in not only management, but also in environmental change, politics, economic behavior, medicine, engineering, and other fields" [5, p. 5].

However, it is not just a flexible approach to the description of business processes in a firm or in an organization but moreover it is a powerful toolset for an enterprise designer. The differential equations laying the mathematical foundation of system dynamics are not a part of everyday toolset of the majority of financiers. In the former USSR, an attempt to apply the methods of analogue simulation (based on the same differential equations) to the research of economical objects was made by V.A. Trapeznikov and his scientific school [15] in the 1960s.

**2. The brief review of present state**

The concept of system dynamics implies a maximum level of the abstraction of a model by describing the structure of a system and system behaviour as a set of interacting positive and negative feedbacks and delays. By operating with processes that are continuous in time, the system dynamics apparatus allows a researcher to represent a simulation model in the form of a system of differential equations (mapping real economical processes to inflows and outflows between Funds).

The methods of system dynamics are implemented in a significant number of software tools (DYNAMO, Vensim, PowerSim, iThink, AnyLogic™ etc.).

The long-term international case record of this technique has clearly demonstrated its advantages compared to competitive methodologies. First of all, system dynamics models are essentially more effective for detecting and understanding of behavior patterns compared to conventional economic-mathematical models. Secondly, they easily allow for corrections in the form of Forrester's delays and nonlinearities, and the latter provide for a more precise description of a state of a modelled process (in particular, it is extremely important for the research of banking institutions and their subsystems). Finally, the approach of system dynamics allows a researcher to estimate even low couplings (when some factors cannot be credibly measured) and to introduce these factors in the model in a way that makes it possible to extend the number of predicted scenarios of behavior of a studied object (by a principle "What will happen, if?...").

It is not hard to explain the attention of a large number of scientists and professionals to the problems of applying system dynamics to economics; operating control, forecasting, and strategic planning all require regular simulation both of regular activity and of critical situations with the goals of the best preparedness to events and the best adaption to the environment.

In the opinion of the author, it is useful to review some contemporary publications directly related to the problems under consideration.

For example, in the article [4], S. Braje has presented an approach to credit risk modeling that builds on standard techniques to develop a system dynamics model. To Braje's knowledge, this is the first attempt to analyze a loan portfolio using transition matrices within a system dynamics framework. The paper shows how a simple model considering the asset and cash flow structure of a loan portfolio can give a valuable information about the performance of the portfolio over time for the further analysis of the steady state equilibrium. Simulations based on models with varying levels of details indicate a significant sensitivity of results to the depth of the dynamic structure representation and to the decision heuristics that are used to determine growth strategies and pricing. The latter holds true even in the simplified framework, in which a single and isolated bank is modeled in an environment of a fixed funding rate, and the response to a single discrete change in parameters is considered.

In his turn, A. Moscardini (with the colleagues) considers the evaluation of new bank loans to be one of the main dilemmas that banks managers have to deal with in order to reduce the probability of default [9]. The lending process is a series of activities involving two main parties whose association ranges from the loan application to the successful or unsuccessful repayment of the loan. This paper describes the construction of a simulator which uses the ideas of system dynamics and the Viable Systems Methodology. The decision support tool thus formed uses systemic approaches to measure a firm's performance and can provide a risk assessment in the sense of evaluating performance under different scenarios. The credit worthiness from this model can then be evaluated against a more traditional estimate based only on financial ratios.

E. Tymoigne's paper [16] presents a model that studies some of the features of Minskyan Framework. The model is Post-Keynesian in nature and puts a large emphasis on the role of conventions and on the importance of the financial side. In doing so, Tymoigne provides an innovative way to determine aggregate investment and to introduce nonlinearities in the modeling of Minsky's framework. This nonlinearity relies on the shifting property of conventions and on the behavioral and psychological assumptions that they carry. Another specific characteristic of the model is that it is stock-flow consistent and explicitly takes into account the amortization of the principal and refinancing loans. In much the same way as in [4] and [9], all of the modeling is done by using system dynamics.

### 3. Formulation

A.R. Gorbunov has remarked: "The activity of a commercial bank consists in the transformation of a flow of the attracted capitals in a flow of bank transactions involving financial assets. According to this general plan, most accessible and straightforward simulation models of banks are developed" [7, p. 174]. It follows that the system dynamics model of a bank should reflect both positive and negative feedbacks being most representative for the financial activity. These feedbacks are linking the following Forrester stocks: {*deposits*; *interbanking receipts*; *other liabilities*} and {*loans*; *investment*; *other assets*} in the course of the conversion of liabilities into assets. It is necessary to take into account that a universal bank belongs to the class of very complex stochastic self-adjusting systems with a great number of feedbacks (according to the classification by S. Beer [3, pp. 33-36, 53-56]). With the course of time, the quantity and complexity of links increase (along with the extension of a bank branches network, the advance of new products on the market of banking services etc.).

On the other hand, a bank is possible to represent as a non-stationary finite-dimensional system with continuous time (in the sense of Kalman [8]) because the real economical processes flow in continuously varied time. Time is one of the parameters of the applicable models; the obvious necessity of the usage of a time factor is determined by the presence of inflation processes and the depreciation of money during a period of time.

Thus, within the framework of Forrester methodology and relying upon the our precursor articles [11-14], purpose of the present research is determined as the construction of a high level model of a bank by means of system dynamics in which the bank is modeled by a set of queuing systems.

### 4. Some foundations and model

In the first place, the nature of a bank's behavior is determined by its info-logical structure reflecting technological aspects of business processes along with the corporate policy and traditions (that directly or indirectly determine the process of decision making). It is applicable to the transformation of financial flows inside a bank first of all.

In consequence of this, a simulation process of the indicated transformation requires the fulfilment of the following steps:
   i)    the design of the bank skeleton diagram, including major sources of amplifications and delays for input-output flows together with information feedbacks;
   ii)   the detection of the basic resource flows (the staff of the bank, money resources as assets and liabilities separately, client applications for bank services, equipment and materials, and also associated information flows);
   iii)  the formalization of administrative communications in the form of the systems of differential equations derived from the use of Forrester's categories {*stock*; *flow*} (in full agreement with the methodology of system dynamics).

Following our papers [11, 12], we shall remark that bank's organizational structure together with delays of the administrative solutions and operations, and also the business rules, can be mapped to the generalized view by means of some algebraic construction, homomorphic to the modelled system relative to the set of predicates defined on the system. For an arbitrary time $t$, the bank can be described by an aggregated index of productivity $D(t)$ that corresponds to the bank state at some point of $k \times n$-dimensional financial space $\mathbf{D}^{k \times n}$ (where $k$ denotes a number of economical indicators for evaluation, and $n$ denotes the quantity of bank branches). Thus, during the present research the main emphasis is made on a quite high level of abstraction, namely the set of state variables (Forrester's stocks) and time will generate the space in which the bank phase trajectory (flows in dynamics) will be realised.

Our reasoning based upon Forrester's thesis that the banking balance is a level in feedback-loop system. Thereby, it is possible to define the aggregated levels describing financial, material, and human resources of a financial institution: staff, clients with their requirements for bank services, assets, liabilities, equipment, bank office buildings, information resources. Each of listed stocks has associated flows, for example: a growth or a reduction of the number of the bank staff; movement of staff inside the bank; oscillations of clients number etc. (see more details in [13-14]). The typical bank financial flows are selected similarly: receipts of the credits from National Bank, other domestic and foreign financial institutions, depository receipts from the individual households, non-cash settlements and money transfers, etc.

To accomplish the evaluation of the dynamical characteristics of a bank in the system sense, we need to known the nature and values of delays for resource flows and administrative influences. These delays are often due to the features of the financial technologies (know-how) and document circulation in a particular bank, and are caused by human factors too (including both bank employees and clients). We assume that the process of granting some unit of the banking product to a client (meta-operation by [11]) consists of $q$ different elementary operations. In its turn, the meta-operation fulfillment cost is a discrete function of a runtime of these operations and can be expressed in a matrix form. The dimension of a matrix $D$ of temporal parameters of the model is $q \times (k_1+k_2)$, where $q$ stands for a number of meta-operations (number of rows in a matrix), $k_1+k_2$ is a quantity of columns in a matrix (each of the columns is responsible for a particular kind of work time costs, $k_1$ represents a quantity of the applicable elementary operations executed by the bank staff, and $k_2$ denotes a quantity of different external delays).

Thus, matrix elements $d_{ij}$ represent the values of the type $j$ of time cost for the type $i$ of a manufacturing process (a meta-operation). In practice, besides delay sources, the sources of amplification in the system with $d_{ij} < 0$ can exist. In the general case, the values of $d_{ij}$ can oscillate about the average points (for example, they are increasing during tax payments).

After all, the total time expenses of manufacturing and of the promotion of a unit of the banking product (taking into account all delays and amplifications) can be expressed in the form:

$$d = \sum_{i=1}^{q} \sum_{j=1}^{k_1+k_2} d_{ij}$$

Likewise, we define a matrix $V$ of cost parameters of the model (a value of the element $v_{ij}$ is the cost of a unit of time for the type $j$ of an elementary operation or an external delay for the type $i$ of a manufacturing process). Then, the production costs of a unit of the banking product (taking into account only the fulfilment operations and unproductive losses of time) can be written as:

$$v = \sum_{i=1}^{q} \sum_{j=1}^{k_1+k_2} d_{ij} \times v_{ij}$$

The elements of a matrix $D$ are associated with substantial delays happening during the rendering of any service to clients of the bank including designs of the credit or depository agreements, the maintenance of clearing accounts, collections of bills, the implementation of interbanking payments, etc.

Having prescribed applicable input and intermediate variables for each stock (Forrester's level) and its flows, we are now in a position to initiate the parameterization of the simulation model of a banking institution. "The system description is translated into the level and rate equations of a system dynamic model" [5]. The rates of flows reflect the dynamics of the change of levels; the data on levels are input values for rate equations. It allows us to formalize the feedback-loop structure of a bank (for more details we refer the reader to [13]). Roughly speaking, assets and deposit income participate in positive feedback loops of profit while liabilities and deposit expenses of bank participate in negative feedback loops.

The isolation of feedbacks loops allows us to describe processes that are happening in the bank management system. The tool being applied for this purpose is systems of ordinary differential equations (see too [10]). It is necessary to point out that having applied this method, we can search for hidden "bottlenecks" in a system more easily – by means of the analysis of a ratio between service rates (intensities) and bank service application rates. In fact, this approach can help us predict the appearance of bottlenecks in the future also; i.e., simulation model conducts to better understanding of banking business processes, and bank staff can receive the answer to a problem "What and why?" at any time and for every bank subsystem.

Note that we can't avoid completely the influence upon bank behavior of the random perturbations of an environment (and bifurcations arising from these perturbations [2]). We use the advantage of phase space non-uniformity that allows us to distinguish areas where system dynamic model is applying almost wholly; for remainder areas, we may accept the probabilistic methods [17]. On the other hand, the sudden stepwise variations of system's regular state at smooth variations of external conditions (in the sense of the catastrophe theory [2]) can been considered manifestations of Forrester's amplifications and long time delays.

In a perfect case, wholly generalized equations of a model should encompass all phases including both the changes of the bank structure and the changes of the mode of the interaction with an environment (first of all, changes of a boundary layer of a system, bank's front-office). For example, it is necessary to perform beforehand simulations of representative banking situations: an excess accumulation of demand deposits and a downsizing of allocation of resources, an appreciation of tariffs rates of main bank transactions (tariffs are up to "the limit of tolerance"), etc. V. Arnold pays attention to the relevance of a similar simulation for the economical applications: "The optimization and intensification can result in a disastrous loss in stability" [2, p. 98].

By the way of illustration, we show a part of the system dynamic model of bank; namely, we attempt to describe **the promotion of a set of banking products among some categories of clients** (in the supposition, that it is a stochastic process with discrete states and continuous time). We make the following assumptions:
  i) in the case of continuous Markov circuits and Poisson events flow, the densities of transition probabilities represent the intensities of events flows;
  ii) if the number of system states is finite (although being a large value), then under the assumption of the persistence of intensity of events flow at $t \to \infty$ some limiting steady-state regime is established [1];
  iii) system components are uniform (within the limits of considered categories of clients).

Thereby, we can take advantage of Kolmogorov equations to formalise a mathematical model [ibid., p. 315]. It is an important step towards the model that naturally combines the discrete view of the states of a bank as a set of queuing systems with the continuous view of the operational worktime of a bank. Along the way, this approach makes it possible to take into account the interactions between the objects of simulation and an input stream of the service applications (for example, the suspension of the service of clients during staff overload, or the increase of result indicators for this product beyond the planned range). Extra advantages of this approach are the relative simplicity of the model (the set of equations is easily introduced directly on the basis of the labelled graph of bank system states), and also the capability of the model to account for both the specialized applications service centers (bank staff) and the operational load.

Let client base consist of a set of $N$ homogeneous elements (clients), $\varepsilon_i$ be a state of an element (the depository agreement is concluded; the deposit income is repaid; deposit and accrued interest are repaid completely; the agreement is prolonged etc.), $\delta$ be the intensity of the update of the state $\varepsilon_0$ (signing of the first agreement for a new client), $m_i$ be the expectation of the number of elements located in a state $\varepsilon_i$, $\lambda_{ij}$ be the intensity of flows of clients for the transition $\varepsilon_i \to \varepsilon_j$. We set the sum of expectations $m_i$ equal to $N$ as a normalizing condition. Then it is possible to represent the system of differential equations in the form:

$$\frac{dm_i}{dt} = -\sum_{k \in K} m_i I_{ik} + \sum_{l \in L} m_l I_{li}$$

In the latter differential equation, $K$ is the set of states for which transitions from $m_i$ are exist; $L$ is the set of states for which transitions into $m_i$ are exist; on the right side of the equation, the value $\delta$ should be included for the state $m_0$. The input data for an initial estimation of the intensities of client flows and expectations can be obtained both from the time records on the activity of staff of a particular bank (for the period of 2 to 4 weeks) and from the applicable operational documentation and databases.

Let's note that in practice the shift of bank's staff attention to yet not "enveloped" clients and the concentration of efforts on them is happening relatively slowly due to the delays of trusted data on the state of the client sector. At the same time, by no means real clients are passive sacrifices of bank institutions; vice versa, they counteract by individually estimating financial companies and by selecting particular banks. Thus, we need to make such a correction of business processes that takes into account the bank resources reallocation and the "resistance" of potential clients.

Having applied the methods of operations research (see [1, p. 322-325]), we can write the equations for interactions in the system {*banks*; *clients*}. For example, expression for banks (for clients analogously):

$$\frac{dm_B(t)}{dt} = -p_C \times \lambda_C(t - \tau_C) \times m_C(t - \tau_C) \times \frac{m_B(t)}{m_B(t - \Delta_B)}$$

Here, $p_B$ and $p_C$ are probabilities of success (attachment of new clients by a bank, and a selection of a new "good" bank by client), $\tau_B$ and $\tau_C$ are operation times for a bank and a client respectively, $\Delta_B$ and $\Delta_C$ are time delays in the course of attention switching for a bank and a client respectively, $\lambda_B(t-\tau_B)$ and $\lambda_C(t-\tau_C)$ are the productivities of the parties at the interaction start time, $m_B(t-\tau_B)$ and $m_C(t-\tau_C)$ are numbers of active interested banks and clients respectively at the interaction start time, $m_B(t-\Delta_B)$ and $m_C(t-\Delta_C)$ are the numbers of active banks and clients respectively at the time of information receipt, $m_B$ and $m_C$ are the numbers of active banks and clients respectively that are not yet in interaction at time *t*.

The following conditions should be added:
1) $m_C(t-\tau_C) = N_C, \quad 0 \leq t \leq \tau_C$
2) $m_B(t-\tau_B) = N_B, \quad 0 \leq t \leq \tau_B$
3) $m_C(t-\Delta_C) = N_C, \quad 0 \leq t \leq \Delta\tau_C$
4) $m_B(t-\Delta_B) = N_B, \quad 0 \leq t \leq \Delta\tau_B$
5) $N_B$ and $N_C$ are primary numbers of banks and clients respectively that are beyond the confines of the domain of partner relations.

Without being too bold, we can state that this approach is still applicable to more abstract schemas of bank service in the system {*deposit owners*; *banks*; *loaners*}. To our regret, it goes beyond the frameworks of the present article and requires a special consideration. It remains to note that the implementation a system dynamic model of a bank in the form of a set of queuing systems (similar to the one described above) using the simulation systems (Vensim, GPSS World, AnyLogic™ etc.) enables the reconfiguring of banking business processes before the development of the negative tendencies.

## 5. Conclusions

Much research activity in the past 50 years has been directed at improving the method of system dynamics for the economic analysis. Under today's conditions of an ongoing economic crisis, those tools that are able to offer a rare combination of speed, usefulness, and affordability of decision making will be of a special value. Based on the realistic consideration of the practical requirements to a skill level of the users of computer expert systems, we believe that one of perspective directions of scientific researches be the creation of simulators supporting everyday activity of the banking personnel. System dynamics provides such an instrument not only for any administrative level but also for managerial decisions and actions of any scale.

The justified selection and skilful usage of the simulation toolkit facilitate the optimization of a bank's organizational structure, banking operations and their technological routes, and supporting information flows. All of these promote the profit growth and decrease the needs in material, financial, and human resources. Besides providing us with the adequate characteristics of the current and future activity of a bank, the application of the system dynamics method allows us to proceed further with the exploration of the management strategies and decisions providing for an extended success of an economic activity (i.e., providing for a desirable value of the effectiveness criterion). The application of these tools by decision makers promotes more precise and timely comprehension of the latent reasons of bank troubles. The system dynamics technique also enables faster and more precise localization of the sources of malfunctions, and a subsequent execution of a calculated pointed liquidation of redundant or non-productive bank branches (or a dosed correction of the business rules).

This article develops the hybrid method of the modeling of banking business processes. The hybrid method under consideration is a combination of the system dynamics philosophy, queueing theory, and Kolmogorov equations apparatus. The paper provides many opportunities for further exploration, and a huge amount of work has to be done. Many directions that have not been explored in this work require further investigation. The approach developed in the paper needs to be applied to concrete simulation scenarios.

We will try to suggest some particular applications of the presented research. This approach may be used to implement the following procedures:

- The optimization of the quantity and of the tasks of banking branches staff;
- The rationalization of a bank's infrastructure (branch network, ATM-network, POS-terminal network etc.);
- The aggregation of the operational information for planning and forecasting;
- Business simulation games for the selection and the training of the candidates for the vacant positions in a bank;
- The training of the students of the applicable profiles using a training "baby" bank;
- The analytical phase during the design or the redesign of banking information systems.

All the procedures above will promote the growth of the bank efficiency both in the tactical and in the strategic sense; after decreasing operational expenses, the subsequent improvements will be gradual steps towards the "personnel-free front-office" pattern (on the basis of intellectual ATMs coupled with other measures).

**Acknowledgments**

The author is grateful to the anonymous referees for their careful reading of the paper and for the suggestions. We would also like to thank Prof. Alexander Gofman (Moskowitz Jacobs Inc.) for his hidden support with our research and for various stimulating and helpful discussions. Finally, I am also grateful to Aleksandr Rafalovich for his numerous helpful remarks that improved the readability of the article.